\documentclass[runningheads,a4paper]{llncs}

\usepackage{amssymb}
\usepackage{graphicx}

\usepackage{url}
\urldef{\mailsa}\path| mahyunst@yahoo.com, samn@ftsm.ukm.my |
\newcommand{\keywords}[1]{\par\addvspace\baselineskip
\noindent\keywordname\enspace\ignorespaces#1}

\begin{document}

\mainmatter  

\title{Keyword Extraction for Identifying Social Actors}

\titlerunning{Keyword Extraction for Identifying Social Actors}

%
%
\author{Mahyuddin K. M. Nasution$^{1,2}$%
\thanks{}%
\and Shahrul Azman Mohd Noah$^{2}$}
\authorrunning{M. K. M. Nasution and S. A. M. Noah}

\institute{$^{1}$Departemen Teknologi Informasi, FASILKOM-TI,\\
Universitas Sumatera Utara, Padang Bulan 20155 USU, Medan, Indonesia.\\
$^{2}$Knowledge Technology Research Group, CAIT-FTSM\\
Universiti Kebangsaan Malaysia, Bangi 43600 UKM Selangor Malaysia.\\
\mailsa\\}

%
%

\toctitle{draf}
\tocauthor{}
\maketitle

\begin{abstract}
Identifying the social actor has become one of tasks in Artificial Intelligence, whereby extracting keyword from Web snippets depend on the use of web is steadily gaining ground in this research. We develop therefore an approach based on overlap principle for utilizing a collection of features in web snippets, where use of keyword will eliminate the un-relevant web pages.
\end{abstract}

\section{Introduction}

One of tasks in extracting information about social actor such as in social network extraction is identification of appropriate actor \cite{nasution2010}. Today, together with information explosion on Internet, it is difficult to associate the web pages to the intended social actor correctly and precisely, mainly by the presence of semantic relation as synonymy and polysemy. For example, in DBLP the author name "Tengku Mohd Tengku Sembok" is sometimes written as "Tengku Mohd Tengku Sembuk", or "Tengku M. T. Sembok", in another case, an actor named of "Shahrul Azman Mohd Noah" has a name label as "Shahrul Azman Noah", are two of many cases about social actor which may have multiple name variations/abbreviations in citation across publication. In other case, names like "Michael D. Williams" and "Mark D. Williams" have used another label name as "MD Williams" in some citations. It is a case of different social actors may share the same name label in multiple citations. 

Most of works have addressed two above problems as a name disambiguation or a co-reference among them are the preparing of person-specific information \cite{mann2003}, the finding the association of person \cite{bekkerman2005}, distinguishing the different persons with keyword/key-phrase \cite{li2005}, domain of research paper citation \cite{han2005}, etc. However, little works attempt to extract flexible features of person as keyword for extracting social networks from Web, or to deal with the information explosion on the Internet that is increasing and expanding the gap of relationship between the actors and words continuously. This paper aims to address keyword for identifying social actors with exploring Web snippets as source of flexible feature. 

\section{Problem Definition}
In semantic the disambiguation is the process of identifying related to essence of word, the sense inherited to objects or social actor names, and also the meaning embedded to it by how societies use it. The meaning found in dictionaries, the dictionaries were written based on events within social, and they were distributed in web pages, while sometimes an event exists in/on the another events. Therefore, the disambiguation issues are subject to kind of overlap principles, i.e. a paradigm to understand some event in the world by the intersection parts, which distinguishes one actor as special case of WSD (word sense disambiguation) \cite{schutze1998}, especially for regarding a person \cite{mccarthy2004}. There are motivations of disambiguation problem, i.e.

\begin{enumerate}
\item Meronymy \cite{nguyen2008a,bollegala2006}:  x is part y or "is-a", part to whole relation - the semantic relation that holds between a part and the whole. In the other word, the page for x belongs to the categories the y. For example, the page for the Barack Obama in Wikipedia belongs to the categories (a) President of United States, (b) United States Senate, (c) Illinois Senate, (d) Black people, etc. In other case, some social actors are associated with one or more categories. For example, Noam Chomsky is a linguist and Noam Chomsky is also a critic of American foreign policy.
\item Holonomy: x has y as a part of itself or "has-a", whole to part relation - the semantic relation that holds between a whole and its parts. For example, in DBLP, the author name "Shahrul Azman Mohd Noah" has a name label as "Shahrul Azman Noah".
\item Hyponymy \cite{nguyen2008b}: x is subordinate of y or "has-property", subordination - the semantic relation of being subordinate or belongs to a lower rank or class. In other word, the page for x has subcategories the y. for example, the homepage of Tengku Mohd Tengku Sembok has categories pages: Home, Biography, Curriculum Vitae, Gallery, Others, Contact, Links, etc. Some pages also contain name label "Tengku Mohd Tengku Sembok"
\item Synonymy \cite{lloyd2005,song2007a,song2007b}: x denotes the same as y, the semantic relation that holds between two words or can (in the context) express the same meaning. This means that names of actor may have multiple label variations/abbreviations in citations across publications. For example, in some web pages, the author name "Tengku Mohd Tengku Sembuk" is sometimes written as "T Mohd T Sembok", "Tengku M T Sembok".
\item Polysemy \cite{song2007a,song2007b}: Lexical ambiguity: individual word or phrase or label that can be used to express two or more different meanings. This means that different actors may share the same name label in multiple citations. For example, both "Guangyu Chen" and "Guilin Chen" are used as the co-reference of "G. Chen" in their citations.
\end{enumerate}

Let $A$ is a set of actors, i.e. $\{a_i|i = 1,\dots,n\}$. We can classify some sets of actors based on the labels of actors.

\begin{enumerate}
\item $A_d = \{a_j|j=1,\dots,m\}$ is a set of ambiguous names which need to be disambiguated, e.g., $\{$"Abdullah Mohd Zin", "Shahrul Azman Mohd Noah", $\dots\}$. Thus, $A$ be a table of reference names containing actors which the names in $A_d$ represent: $A = \{$"Abdullah Mohd Zin (academic)", "Abdullah Mohd Zin (policitian)",$\dots\}$.
\item $A_t = \{a_l|l=1,\dots,k\}$ is a set of composition of names token (first/middle/last name or abbreviation), and due to an actor has multiple name variations: "Shahrul Azman Mohd Noah (Professor)" may appear in multiple web pages under different names such as "Noah, Shahrul Azman Mohd", "Shahrul Azman Noah", "Shahrul A. Noah", "Noah, Shahrul A.", "Shahrul Azman MN", and Shahrul Azman bin Mohd Noah", where "bin" is special token meaning "son of".
\item $A_x = \{a_p|p=1,\dots,o\}$ is a set of actors which can be observed in web pages (documents). The names need patterned and disambiguated. The actor's names can be rendered differently in online information sources. They are not named with a single pattern of tokens only, they are not also labeled with unique identifiers, and so they are uncertain persons.
\end{enumerate}

Therefore, name disambiguation is an important problem in extracting information of social actors whereby the social actors can be expressed by using different aliases due to multiple reasons as motives: use of abbreviations, different naming conventions, misspelling, pseudonyms in publication or bibliographies (citations), or naming variations over time. We conclude that there are two fundamental reasons of name disambiguation in semantic for identifying actors, as follows:

\begin{enumerate}
\item There are a relation $\phi$ to assign $\Omega$ as space of events containing actors to $A$: a relation $\phi_1 : \Omega\stackrel{1:M}{\longrightarrow} A_x$ and a relation $\phi_2 :A_x\stackrel{N_1+N_2}{\longrightarrow} A$, $M = N_1+N_2$, where $\phi_1$ and $\phi_2$ are relations $\phi_t : A_t \stackrel{N_1:1}{\longrightarrow} A$ and $\phi_d : A_d \stackrel{N_2:1}{\longrightarrow} A$. $A_t, A_d \subseteq A_x$  such that $\phi = \phi_1\phi_2 : \Omega\longrightarrow A$.
\item There are a relation $\varphi$ to assign $A$ containing $a$ to $\Omega$, i.e. there are a relation $\varphi_t : A \stackrel{1:M_1}{ \longrightarrow} A_t$ and a relation $\varphi_d : A\stackrel{1:M_2}{\longrightarrow} A_d$. $A_t, A_d \subseteq A_x$ for getting a relation $\varphi_1 = \varphi_t\varphi_d : A \stackrel{1:M_1+M_2}{\longrightarrow} A_x$ where there is a relation $\varphi_2 : A_x\stackrel{M_1+M_2:N}{\longrightarrow} L$, $L\subseteq\Omega$ such that $\varphi = \varphi_1\varphi_2 : A \longrightarrow L$.
\end{enumerate}

\section{The Proposed Approach}
The start of this approach is some concepts as follows.

\begin{enumerate}
\item A word $w$ is the basic unit of discrete data, defined to be an item from a vocabulary indexed by $\{1,\dots,K\}$, where $w_k = 1$ if $k \in K$ and $w_k = 0$ otherwise.
\item A term $t_k$ consist of at least one word, or $t_k = \{w_1,\dots,w_l\}$, $l \le k$, $k$ is a number of parameters representing word, and $|t_k| = k$ is size of $t_k$. 
\item Let a web page denoted by $\omega$ and a set of web pages indexed by search engine be $\Omega$ containing pairs of term and web page. Let $t_x$ is search term and a web page contains $t_x$ is $\omega_k$, we obtain $\Omega_x = \{(t_x,\omega_x)\}$, $\Omega_x\subseteq\Omega$, or $t_x\in \omega_x \in \Omega_x$. $|\Omega_x|$ is a cardinality of $\Omega_x$.
\item Let $t_x$ is a search term. $S = \{w_j,\dots,w_{max}\}$ is a web snippet (briefly snippet) about $t_x$ that returned by search engine, where $max = \pm 50$ words. $L = \{S_i|i = 1,\dots,m\}$ is a list of snippet. 
\end{enumerate}

We develop an approach for extracting the keyword from web snippets based on a concept of overlap. A concept that can interpret our world as a composition of the parts and the parts be from other parts, whereby implementation of the overlap principle as intersection operator or logically $t_a$ AND $t_b$ is TRUE in space of event $\Omega$ with two conditions: 
\begin{enumerate}
\item [(a)] Let $t_a$ and $t_b$ are search terms, $t_a\cap t_b$ and web pages contain $t_a$ AND $t_b$ is $\omega_a$ and $\omega_b$, $\omega_a,\omega_b\in\Omega_a\cap\Omega_b$, $|\Omega_a\cap\Omega_b|$ is a cardinality of $\Omega_a\cap\Omega_b$, and $\Omega_a\cap\Omega_b\subseteq\Omega$.
\item [(b)] Let $t_a, tx, ty \in S$ with $|\Omega_x|\le|\Omega_y|$ then $|\Omega_a\cap\Omega_x| > |\Omega_a\cap\Omega_y|$.
\end{enumerate}
The implementation of overlap principle also can use the query of two objects: For each query $query(t_a,t_b)$ submit to search engine, the search engine return a collection of web snippets as a representation of objects in a group of web pages.

Let us model the bag of words (BoW) of snippets to utilize weights of TF.IDF (term frequency-inverse document frequency) \cite{nasution2011a}: TF.IDF $= tf(t_x,S).idf(t_x)$, $tf(t_x,S) = \sum_{j=1,\dots,N}\sum_{i=1,\dots,m} 1/max$, $idf(t_x) = \log N/df(t_x)$ where $max$ is the number of terms/words in a snippet, $m$ is the number of same terms/words in a snippet, $N$ is the number of snippets containing name of actor, and $df(t_x)$ is the number of snippets for term $t_x$ appears. Then, we product a vector space by $v = TF.IDF/hs \in [0,1]$, $hs$ is highest score of TF.IDF, such that $v_i \in [0,1]$, $i = 1,\dots,n$, and $v_1\ge v_2 \ge \dots v_n$. Thus, there is an one-to-one function $\eta = {\bf w} \rightarrow \{v_i|i = 1,\dots,n\}$, for ${\bf w}$ as a set of vocabularies. A micro cluster of words ${\bf w}$ \cite{nasution2010} is words network $G_w$ from Web by using Jaccard coefficient to hit counts such that
\[
sim_{jac}(t_x,t_y) = \frac{|\Omega_x\cap\Omega_y|}{|\Omega_x|+|\Omega_y|-|\Omega_x\cap\Omega_y|},
\]
and the optimal micro cluster $T_w$ is words tree, it is used for eliminating the irrelevant word from list of candidates or it can represent the same actor. Thus there is a subset of words ${\bf w}_o = \{w_i|i=1,2,\dots,m\}$, $m\le n$, and wo is subset of ${\bf w}$. However, any words in tree still wo be the characteristic of two and more actors. We model a vector space of words tree $u_i$ by dividing all their hit counts by highest hit count relatively: there is $\rho : {\bf w}_o\rightarrow \{u_i|i=1,2,\dots,m\}$. Based on the conditions (a) and (b) of overlap principle we define $\delta_i = v_i - u_i \ge 0$ for selecting a keyword $t_x$ from the optimal micro cluster, i.e. for $\delta_x > \dots > \delta_y$, $t_x$ is keyword. The steps for extracting keywords are as follows.

\begin{table}
\caption{An optimal micro cluster of an actor: "Abdullah Mohd Zin"}
\label{tabel:kata}
\begin{center}
\begin{tabular}{lccc}\hline
Words         & $v$     & $u$     & $\delta$\cr\hline
network       & 1.00000 & 0.32680 & 0.67320\cr
international & 0.57339	& 0.49485 & 0.07855\cr
computer      & 0.56474 & 0.48814 & 0.07660\cr
system        & 0.52506 & 0.52577 &-0.00072\cr
software      & 0.50420 & 0.68041 &-0.17621\cr
use           & 0.40142 & 1.00000 &-0.59858\cr\hline
\end{tabular}
\end{center}
\end{table}

\noindent 
Generate(keyword)\\
INPUT  : A set of actor\\
OUTPUT : keyword(s) of each actor\\
STEPS  :\\
\begin{enumerate}
\item $w = |\{w_i|i=1,2,\dots,n\} \leftarrow$ Collect terms/words per actor from snippets.
\item $\{v_i|i=1,2,\dots,n\} \leftarrow$ Generate vector for all $w\in {\bf w}$ by using TF.IDF.
\item $\{u_i|i=1,2,\dots,n\} \leftarrow$ Generate vector for each hit count of $w \in {\bf w}$.
\item $G_w \leftarrow$ Build the micro cluster using hit counts.
\item $T_w \leftarrow$ Make the optimal micro cluster based on $G_w$.
\item If $T_w$ do not consist of trees, then collect and cut node with degree $deg > 1$ for separating $T_w$ be trees.
\item Select a cluster from trees of $T_w$ by using a predefined stable attribute.
\item Find maximum $\delta$ from candidate keywords in a cluster for generating the keyword.
\end{enumerate}

\section{Experiment}

Let us consider information context of actors that includes all relevant relationships with their interaction history, where Yahoo! search engines fall short of utilizing any specific information, especially micro cluster information, and just therefore we use full text index search in web snippets. In experiment, we use maximum of 500 web snippets for search term $t_a$ representing an actor, and we consider words where the TF.IDF value $> 0.3\times$ highest value of TF.IDF, or maximum number is 30 words. $w = \{$network, minister, Malaysia, journal, datuk, department, Allah, international, Ismail, Nazri, computer, prime, ictac, learning, system, software, foxley, said, kebangsaan, performance, dr, university, Eric, use, accuracy, dblp, based, communications, utilization, author$\}$ for example is a set of 30 words from web snippets for actor "Abdullah Mohd Zin".  We test for 143 names, and we obtain 8 ($5.59\%$) actors without a cluster of candidate words, 13 ($9.09\%$) actors with only one cluster, and 122 ($85.32\%$) persons have two or more keywords. In a case of "Abdullah Mohd Zin" we have $\{$network, international, computer, system, software, use$\}$, $\{$Malaysia, accuracy$\}$, $\{$datuk, Nazri, kebangsaan$\}$, $\{$minister, journal, ictac, dblp, communications, utilization$\}$, $\{$department, learning, said, performance$\}$, $\{$dr, university, based$\}$, $\{$prime, foxley, Eric, author$\}$, $\{$Allah, Ismail$\}$ as micro clusters of words. We can arrange the individual keywords according to their proximity to the stable attribute "academic", i.e. a set of words in SK $= \{$sciences, faculty, associate, economic, prof, environment, career, journal, network, university, report, relationship, context, $\dots\}$. SK and $\delta$ maximum exactly determine that "network" be a keyword for actor "Abdullah Mohd Zin" as an academic (not a politician), in Table \ref{tabel:kata}. Therefore, we can redefine $\phi$ or $\varphi$ as $query(t_a,t_x)$.

Under the average of recall, precision and F-measure, this method shows something to consider, i.e. the number of words in the cluster should be limited so that so average value of measurement is not affected by the lower, see Table 2. 

\begin{table}
\caption{Average of optimal micro cluster results}
\begin{center}
\label{tabel:avg}
\begin{tabular}{lccc}\hline
Method & Recall & Precision & F-measure\cr\hline
Delta ($\delta$) & $45.8\%$ & $29.5\%$ & $35.9\%$ \cr\hline
\end{tabular}
\end{center}
\end{table}

Some keywords product high recall and low precision, and also the pairs of actor-double-keyword products low recall and precision, because web pages about an actor have greater variety. 

\section{Conclusion and Future Work}

This article presents a practical methodology for extracting social network based on superficial methods (in unsupervised research strem), quality measures for clustering the actors and relations between them with keywords, singleton and doubleton, and URLs address, we obtain the scale $\le n(n-1)/2$. The social network extraction process relies on a dynamic knowledge in Web, thus this is still under development, and research work must be completed and enriched. In particular, a careful next study of measures for their own and compared to each other of superficial methods, mainly to consider complexity and number of submitted queries.

\end{document}